\documentclass[%
 reprint,hidelinks,nofootinbib,
 amsmath,amssymb,
 aps,
]{revtex4-2}

\usepackage{graphicx}
\usepackage{dcolumn}
\usepackage{bm}
\usepackage{mathtools}
\usepackage{csquotes}
\usepackage{hyperref}
\usepackage{float}
\usepackage{bigints}
\usepackage{color}

\begin{document}


\newcommand{\physrep}{{Phys. Rep.}}

\newcommand{\AAA}{\boldsymbol{A}}
\newcommand{\BB}{\boldsymbol{B}}
\newcommand{\JJ}{\boldsymbol{J}}
\newcommand{\EE}{\boldsymbol{E}}
\newcommand{\UU}{\boldsymbol{U}}
\newcommand{\kk}{\boldsymbol{k}}
\newcommand{\xx}{\boldsymbol{x}}
\newcommand{\Bnabla}{\boldsymbol{\nabla}}
\newcommand{\ii}{\mathrm{i}}
\newcommand{\bra}[1]{\langle #1\rangle}
\newcommand{\bbra}[1]{\left\langle #1\right\rangle}
\newcommand{\eqss}[2]{(\ref{#1})--(\ref{#2})}
\newcommand{\EEq}[1]{Equation~(\ref{#1})}
\newcommand{\Eq}[1]{Eq.~(\ref{#1})}
\newcommand{\Eqs}[2]{Eqs.~(\ref{#1}) and~(\ref{#2})}
\newcommand{\Eqss}[2]{Eqs.~(\ref{#1})--(\ref{#2})}
\def\EM{E_{\rm M}}
\def\EA{E_{\rm A}}
\def\HM{H_{\rm M}}
\def\Lu{\mbox{\rm Lu}}
\def \clara#1{{\color{red}CD: #1}}
\def \jose#1{{\color{blue}JP: #1}}

\title{On the Origin of Magnetar Fields: Chiral Magnetic Instability in Neutron Star Crusts}

\author{Clara Dehman$^{1}$}
\email{clara.dehman@ua.es}
\author{Jos\'e A. Pons$^{1}$}

\affiliation{$^{1}$ Departament de Física Aplicada, Universitat d'Alacant, 03690 Alicante, Spain}


\newcommand{\jcap}{{JCAP}}
\newcommand{\aap}{{A\&A}}
\newcommand{\mnras}{{MNRAS}}
\newcommand{\apjl}{{ApJL,}}
\newcommand{\apjs}{{ApJS,}}
\newcommand{\ssr}{Space Sci. Rev.,}

\begin{abstract}

We investigate the chiral magnetic instability in the crust of a neutron star as a potential mechanism for amplifying magnetic fields. This instability may become active when small deviations from chemical equilibrium are sustained over decades, driven by the star's gradual spin-down or residual heat loss. Our findings suggest that this mechanism can produce strong, large-scale magnetic fields consistent with models that align with observational data. Additionally, this instability naturally generates magnetic helicity in the star's crust, which is crucial for forming and maintaining strong dipolar toroidal fields, often invoked to explain magnetar observational phenomena. 
Our results offer a microphysically-based alternative to classical hydrodynamical dynamos for the origin of magnetar magnetic fields, addressing a long-standing debate in the field.

\end{abstract}

\maketitle

The origin and evolution of neutron star (NS) magnetic fields, particularly in magnetars, have been subjects of long-standing debate  \cite{Mereghetti2015,esposito2021}. Magnetars are known to have the strongest magnetic fields among NSs. It is widely accepted that the fossil field inherited from the progenitor star cannot account for the strongest large-scale dipolar fields observed (inferred from the spin period and its derivative). Thus, amplification by a turbulent dynamo is often proposed as the mechanism to explain such intense fields \cite{balbus1991,obergaulinger2014,raynaud2020,reboul2021,aloy2021}. However, despite numerous studies, certain issues remain unresolved, particularly the energy transfer to larger scales. The origin of magnetar magnetic fields continues to be an open problem.

Recently, new efforts have been made to connect the formation of large-scale magnetic fields in magnetars to a microscopic mechanism: the chiral asymmetry of particles produced during core-collapse supernovae and the proto-NS phase \cite{ohnishi2014,matsumoto2022}. Several challenges have emerged, with the most significant being the efficiency of spin-flip scattering processes in rebalancing the population of left-handed electrons, which are initially depleted when the high electron fraction of the progenitor is converted into left-handed neutrinos during the proto-NS stage. This temperature-dependent effect reduces the chiral imbalance, causing the chiral instability mechanism to fall short in adequately explaining the formation of magnetars' large-scale dipolar fields when all factors are consistently considered \cite{grabowska2015,sigl2016,kaplan2017}.

In this work, we revisit the chiral instability scenario once the NS is cold enough to become transparent to neutrinos and its crust has formed. At this evolutionary stage, NSs are expected to be fully deleptonized and in chemical equilibrium. Denoting the chemical potentials by $\mu_e$, $\mu_\nu$, $\mu_n$, and $\mu_p$, deviations from chemical equilibrium are quantified by the chemical potential difference $\delta \mu = \mu_p + \mu_e - \mu_n$. Here, we neglect the neutrino chemical potential since we are interested in the stages after the NS becomes transparent to neutrinos. 
Several processes studied in the literature can lead to a small deviation from chemical equilibrium throughout the NS's lifespan.
Typically, they arise from the concept that as the star contracts slightly—due to the loss of residual thermal energy, angular momentum, or magnetic field dissipation—the density of the fluid elements within it increases, thereby altering its chemical equilibrium.
Since relaxation to a new equilibrium takes a finite time, matter may not be in perfect chemical equilibrium. Perhaps the most popular scenario involves pulsar spin-down \cite{reisenegger1995,fernandez2005}; as the star spins down, its centrifugal force gradually decreases, leading to contraction and increased fluid density, which disturbs the chemical equilibrium. Accretion can also lead to a similar scenario \cite{wang2017}. It has been proposed that an accretion rate of $\dot{m} \approx 0.2 ~\dot{m}_{Edd}$ is sufficient to compress the URCA shell on a timescale shorter than the weak interaction timescale, causing a departure from chemical equilibrium. Additionally, slow hydrodynamic flows within NS interiors \cite{urpin1996} can cause deviations from equilibrium since the relaxation time for a moving fluid element to reach the new equilibrium is finite. Millisecond oscillations \cite{reisenegger1992,villain2005,villain2005b}, particularly in fast-rotating NSs, can significantly alter their chemical composition by generating centrifugal forces that influence the distribution and migration of elements within the star. These oscillations can also cause plastic flow and cracking in the star's crust, affecting neutrino emissions and leading to complex changes in the star's chemical makeup over time.

Without delving into all these scenarios in detail, we aim to examine the generation of a magnetic field when the NS is slightly out of chemical equilibrium. This deviation leads to a small but persistent imbalance between left- and right-handed electrons, which induces an electric current parallel to the magnetic field, known as the Adler-Bell-Jackiw anomaly \cite{adler1969,bell1969}. The focus of this letter is to investigate this microphysically-based mechanism as an alternative to classical hydrodynamically-based dynamos to generate strong magnetic fields in magnetars.

We denote by $\mu_5 \equiv (\mu_L - \mu_R)/2$\footnote{We follow the sign convention of Ref.~\cite{sigl2016} to define $\mu_5$. Other works \cite{grabowska2015,kaplan2017,rogachevskii2017} adopted the opposite sign convention.} the chemical potential associated with the chiral charge density, where $\mu_L$ and $\mu_R$ are the chemical potentials of the left-handed and right-handed electrons, respectively.
When a chiral imbalance is present ($\mu_5 \neq 0$), Maxwell's equations are modified to include an additional current contribution \cite{ohnishi2014,rogachevskii2017,brandenburg2023}:
\begin{equation}
    \JJ_5 = - \frac{2\alpha \mu_5}{\pi \hbar c} \BB.
    \label{eq: J5}
\end{equation}
Here, $\alpha = e^2/\hbar c$ is the fine structure constant, $e$ is the fundamental charge, $\hbar$ is the reduced Planck constant, and $c$ is the speed of light. This additional current acts as a dynamo term, amplifying the magnetic field at the cost of the chemical energy stored in the chiral imbalance. 

We consider the weak interaction rates in the NS crust, which is composed of nuclei, a gas of ultra-relativistic electrons, and free neutrons in the inner crust. Although the crust already has a low electron fraction, the stratified, neutron-rich nuclei forming the crust still undergo weak reactions of the form
\begin{eqnarray}
    e_L + A  \rightarrow A' + \nu_L ~, \label{eq: e+A} \\
    A \rightarrow A' + e_L + \bar{\nu}_R ~. \label{eq: A decay}
    \label{eq: weak rates nuclei}
\end{eqnarray}
$A$ and $A'$ represent the parent and daughter nuclei, respectively.
In principle, in a perfectly stationary situation, the matter is in exact $\beta-$equilibrium, and the electron capture and its inverse process cancel each other (detailed balance principle). However, during pulsar spin-down or other plausible scenarios mentioned above, residual contraction can still induce some reactions.

On the other hand, because the crust consists of a solid lattice of ions, only the electrons have limited mobility, experiencing a very slow drift. This scenario is known as the Hall-MHD (or e-MHD) limit (see \cite{pons2019} for a review). The evolution of the magnetic field is governed by the induction equation, as described by Faraday's law.
\begin{equation}
    \frac{\partial \BB}{\partial t} = - c \Bnabla \times \EE,
    \label{eq: faraday law}
\end{equation}
where $\EE$, derived from Ohm's law, will now consist of the usual Ohmic and Hall terms plus the new chiral magnetic term:
\begin{equation}
    \EE =  \eta \left(\Bnabla \times \BB - \frac{2 \alpha \mu_5}{\pi \hbar c} \BB \right) + f_h (\Bnabla \times \BB)\times \BB.
    \label{eq: electric field}
\end{equation}
Here, $\eta = c^2/4\pi \sigma$ is the magnetic diffusivity, and $f_h= c/4 \pi e n_e$ is the hall prefactor, where $n_e$ is the electron number density.

The evolution of the magnetic field must be supplemented by the evolution of the chiral number density $n_5$:
\begin{equation}
    \frac{\partial n_5}{\partial t} =  \frac{4 \alpha }{\pi \hbar c} \EE \cdot \BB + n_e \Gamma_w^\mathrm{eff} - n_5 \Gamma_f. 
    \label{eq: n5}
\end{equation}
Here, $\Gamma_w^\mathrm{eff}$ represents the effective weak reaction rate due to deviations from chemical equilibrium and acts as a source term, while $\Gamma_f$ denotes the spin-flip interactions resulting from the finite electron mass, which tend to reduce the asymmetry between left- and right-chiral electrons. The term $\EE \cdot \BB$ transforms chiral asymmetry into magnetic energy. Note the negative sign in front of the chiral term in \EEq{eq: electric field}, which results in a negative $\EE \cdot \BB$ term in \EEq{eq: n5} for a positive $\mu_5$. It is important to mention that the rates of weak reactions and spin-flip interactions have a weak dependence on temperature due to the degeneracy of NS matter. 

In the quasi-steady state, assuming that all quantities except $n_5$ remain constant over time, this equation can be integrated analytically. For times $t \gg \Gamma_f^{-1}$ (which is always true on astrophysical timescales), the following quasi-equilibrium expression for $n_5(t)$ can be obtained:
\begin{equation}
    n_5(t) \approx  
    \frac{4 \alpha }{\pi \hbar c} \frac{\EE \cdot \BB}{\Gamma_f} + n_e \frac{\Gamma_w^\mathrm{eff}}{\Gamma_f}.
    \label{eq: n5(t) analytical}
\end{equation}
 The last term on the right-hand side of \EEq{eq: n5(t) analytical} is the chiral number density at equilibrium in the weak magnetic field limit. The chiral chemical potential $\mu_5$ and the chiral wavelength (or wavenumber) $\lambda_5$ ($k_5$) can be calculated from $n_5$ as follows \cite{sigl2016}:
 \begin{equation}
    \mu_5  = \frac{3 \pi^2 (\hbar c)^3 n_5 }{3 \mu_e^2 + \pi^2 T^2}, \quad\quad  \lambda_5 = \frac{2 \pi^2 \hbar c}{\alpha \mu_5}. 
    \label{eq: mu5 from n5}
\end{equation} 

\begin{figure}
    \centering
\includegraphics[width=\columnwidth]{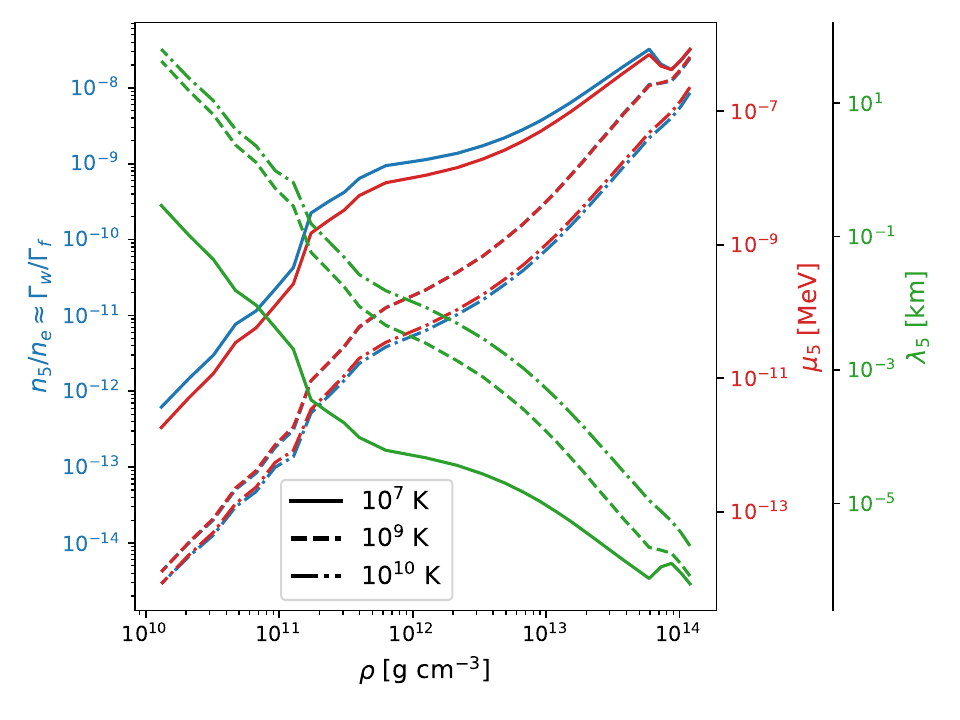}
    \caption{The left y-axis (blue) shows $\Gamma_w/\Gamma_f$, which is equivalent to $n_5/n_e$ in the steady-equilibrium state. The right y-axis (red) shows $\mu_5$, and the secondary right y-axis (green) shows $\lambda_5$, all as functions of density $\rho$. Solid lines represent $10^7$\,K, dashed lines represent $10^9$\,K, and dash-dotted lines represent $10^{10}$\,K.}
    \label{fig: ratio rates and mu5}
\end{figure}

Let us discuss in more detail the reaction rates. Scattered electrons of a certain chirality can flip into the opposite chirality state through Rutherford scattering, electron-electron scattering, or Compton scattering processes, thus decreasing the chiral number density $n_5$. In the core, Rutherford scattering is the dominant process (see, e.g., \cite{dvornikov2015,sigl2016}), whereas in the crust, the main contribution is electron scattering off nuclei. For degenerate electrons, this rate can be expressed as \cite{haensel2007}:
\begin{equation}
\Gamma_f = \bigg(\frac{m_e}{\mu_e} \bigg)^2 \nu_\mathrm{coll} ,  
\label{eq: gamma flip}
\end{equation}
where $\nu_\mathrm{coll} = w_p^2/\sigma$ is the collision frequency, $\sigma$ is the electric conductivity, and $w_p = \mu_e \sqrt{4 \alpha/(3\pi)}$ is the electron plasma frequency. The temperature dependence of electrical conductivity $\sigma$ leads to the flip rate $\Gamma_f$ also being temperature dependent.

The other key ingredient is $\Gamma_w$. The electron capture rate for \EEq{eq: e+A} has been calculated, for instance, in \cite{epstein1981}, assuming a single dominant transition from the ground state of the parent nucleus to a single excited state of the daughter nucleus. Keeping only the leading order terms:
\begin{equation}
\Gamma_w \approx \frac{12 G_w^2 g_A^2}{14 \pi^3 \hbar} \, (Z-20) \, I_\mathrm{fermi} , 
\label{eq: electron capture rate on nuclei}
\end{equation}
where $I_\mathrm{Fermi}$ is the Fermi integral, given by: 
\begin{equation}
    I_\mathrm{Fermi} \approx \mu_e^2 (k_b T)^3 \bigg[ \frac{1}{3} \bigg(\frac{\mu_e - w}{k_b T}\bigg)^3 + \frac{\pi^2}{3}\frac{\mu_e - w}{k_b T}  \bigg].
    \label{eq: fermi integral}
\end{equation}
Here, $G_w \approx 1.166 \times 10^{-5}$\,GeV$^{-2}$ is the weak coupling constant, $g_A \approx 1.25$ is the axial vector coupling constant, and $w \approx m_n - m_p$ is the threshold energy for electron capture, with $m_n$ and $m_p$ being the neutron and proton masses, respectively. 
Notice that, due to the high degeneracy of electrons in the crust of a NS, $\Gamma_w \propto \mu_e^5$, and \EEq{eq: electron capture rate on nuclei} is almost temperature independent. 

In Fig.~\ref{fig: ratio rates and mu5}, we show the ratio of reaction rates $\Gamma_w/\Gamma_f$ (blue), the chiral chemical potential $\mu_5$ (red), and the chiral wavelength $\lambda_5$ (green) as functions of the density in the crust of a typical NS at temperatures of $10^7$, $10^9$, and $10^{10}$\,K. These quantities exhibit a significantly greater dependence on density compared to temperature. It is important to note that both $n_5/n_e \approx \Gamma_w/\Gamma_f$ and $\mu_5$ are very small. Unlike some chiral instability studies in proto-NS evolution models, our scenario does not require $n_5/n_e$ values of order one and $\mu_5$ of order MeV.
This is because the CMI can act over a longer period (tens or hundreds of years) compared to a proto-NS scenario (tens of seconds). For our purposes, a chiral imbalance involving only a tiny fraction of the electrons (less than one in a billion left-handed electrons over right-handed electrons), or equivalently, a $\mu_5$ in the range of $10^{-12}-10^{-6}$\,MeV, is sufficient.
Additionally, the predicted wavelength of the fastest-growing mode is on the order of kilometers near the star's surface, decreasing to centimeters in the inner crustal layers. The smaller the scale, the faster the growth of the CMI \cite{kaplan2017}.  
By the principle of detailed balance, the rate of the inverse reaction (\Eq{eq: A decay}) is given by an identical expression multiplied by a factor $\mathrm{exp}(-\delta \mu /k_b T)$ \cite{epstein1981}, where $k_b$ is the Boltzmann constant and $T$ is the temperature. Therefore, the effective net weak interaction rate in the crust of a NS is: 
\begin{equation}
\Gamma_w^\mathrm{eff} =  \Gamma_w (1 - \mathrm{exp}(-\delta \mu /k_b T)). 
\end{equation}
In exact $\beta$-equilibrium ($\delta \mu = 0$), we have $\Gamma_w^\mathrm{eff}=0$. On the contrary, if slight deviations from $\beta$-equilibrium are allowed,
there could be a small but non-vanishing reaction rate producing a chirality imbalance. In what follows, we assume a very small deviation from chemical equilibrium, of the order of $\delta \mu / k_b T\approx 10^{-2}$. 

\begin{figure}
    \centering
\includegraphics[width=\columnwidth]{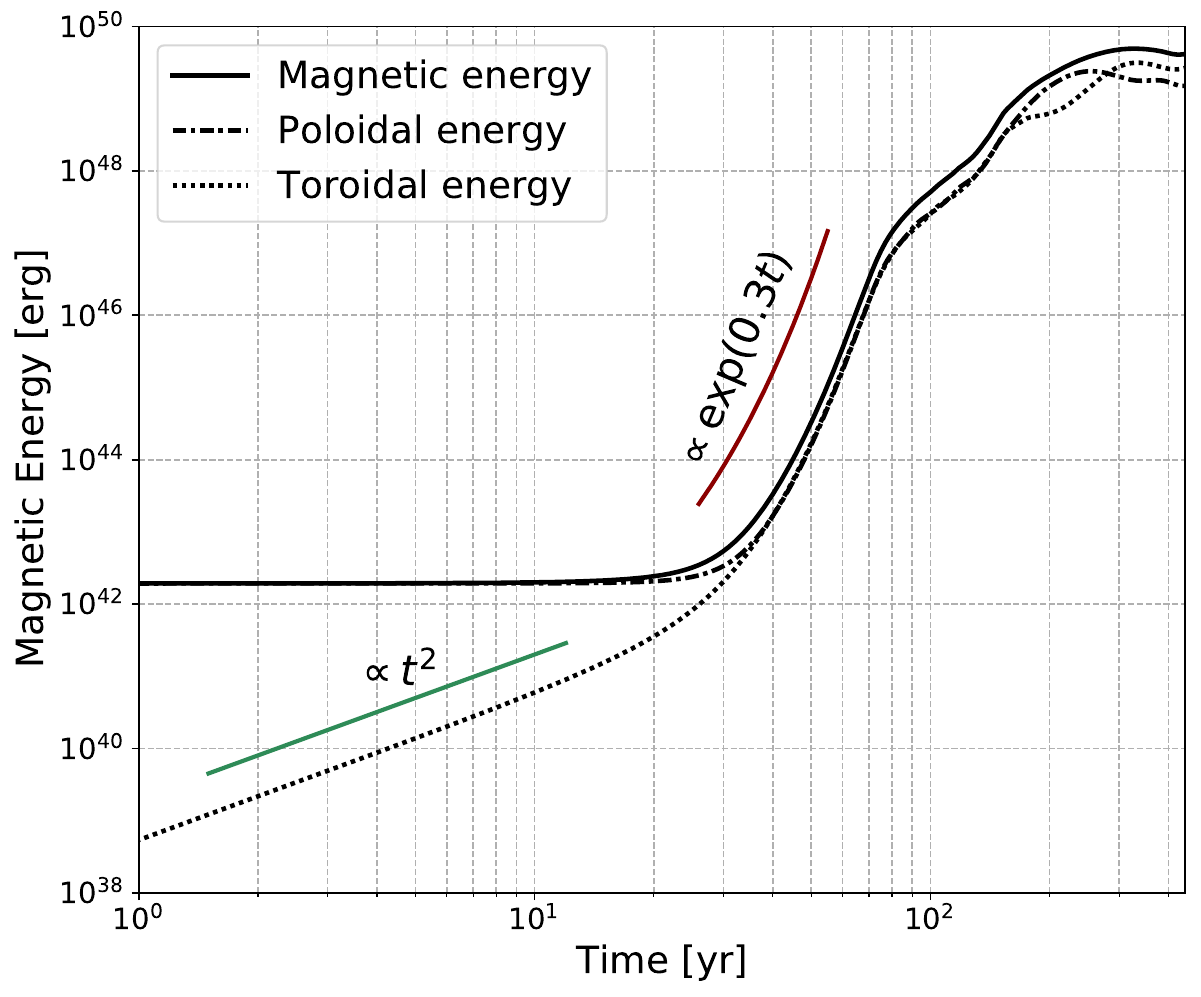}
    \caption{Results from simulations: evolution of the poloidal and toroidal magnetic energy with time. The dash-dotted line represents poloidal magnetic energy, the dotted line indicates toroidal energy, and the solid line shows total magnetic energy. 
    }
    \label{fig: poloidal toroidal energy}
\end{figure}
 
To investigate if the CMI can lead to significant magnetic field amplification in a NS crust, we conducted magneto-thermal evolution simulations using an extended version of the 3D finite-volume \texttt{MATINS} code. This code solves the coupled system of the generalized induction equation and the heat diffusion equation \cite{dehman2023,dehman2023b,ascenzi2024}. We consider an initial magnetic field confined to the crust and employ potential magnetic boundary conditions (current-free magnetosphere) at the outer numerical boundary, placed at a density of $\rho=10^{10}$\,g~cm$^{-3}$, which is close to the transition between the liquid envelope and solid crust in young or middle-aged NSs. At the crust-core interface, we impose perfect conductor boundary conditions. 

We implement state-of-the-art calculations for temperature-dependent electrical conductivity at each point of the star using Potekhin's public codes\footnote{\url{http://www.ioffe.ru/astro/conduct/}} \cite{potekhin2015}. The background NS model can be constructed using various zero-temperature equations of state (EOS) from the CompOSE online database\footnote{\url{https://compose.obspm.fr/}}. Specifically, we present results using the BSK24 EOS \cite{pearson2018} for a star with a mass of $M=1.4 M_{\tt sun}$.

As illustrated in Fig.\,\ref{fig: ratio rates and mu5}, the predicted wavelengths of the growing modes vary significantly, ranging from kilometer scale in the outermost layers of the crust to centimeter scale in the innermost layers of the NS crust. Resolving centimeter scales requires an exceptionally fine radial grid resolution, involving tens of thousands of radial grid cells. Achieving such a resolution is challenging, even for parallelized MHD codes. With a radial grid resolution of around 100 cells, the smallest resolvable scale is a few meters. Given that \texttt{MATINS} is optimized but not parallelized, we consider a grid resolution of $N_r=40$ and $N_\xi=N_\eta=43$ per patch, using a cubed-sphere consisting of six patches \cite{dehman2023}. This configuration corresponds to approximately $l_\mathrm{max} \sim 60$ resolved multipoles. With this grid resolution, we can resolve scales of the order of a few tens of meters. Therefore, we have imposed a cutoff on $\mu_5 \leq \mu_5^\mathrm{max} \equiv 10^{-10}$\,MeV to be compatible with the possible resolved grid resolution. This corresponds to $\lambda_5 \geq \lambda_5^\mathrm{min} \equiv 10^{-2}$\,km ($k_5^\mathrm{max} \approx 600$\,km$^{-1}$). We acknowledge that this is a conservative choice, as we are truncating the smallest-scale, fastest-growing modes in the inner crust for numerical reasons. In reality, these centimeter-scale modes would develop over much shorter timescales than discussed here, although they would also reach saturation and dissipate Ohmically more quickly than the long-scale modes.

\begin{figure}
    \centering
\includegraphics[width=\columnwidth]{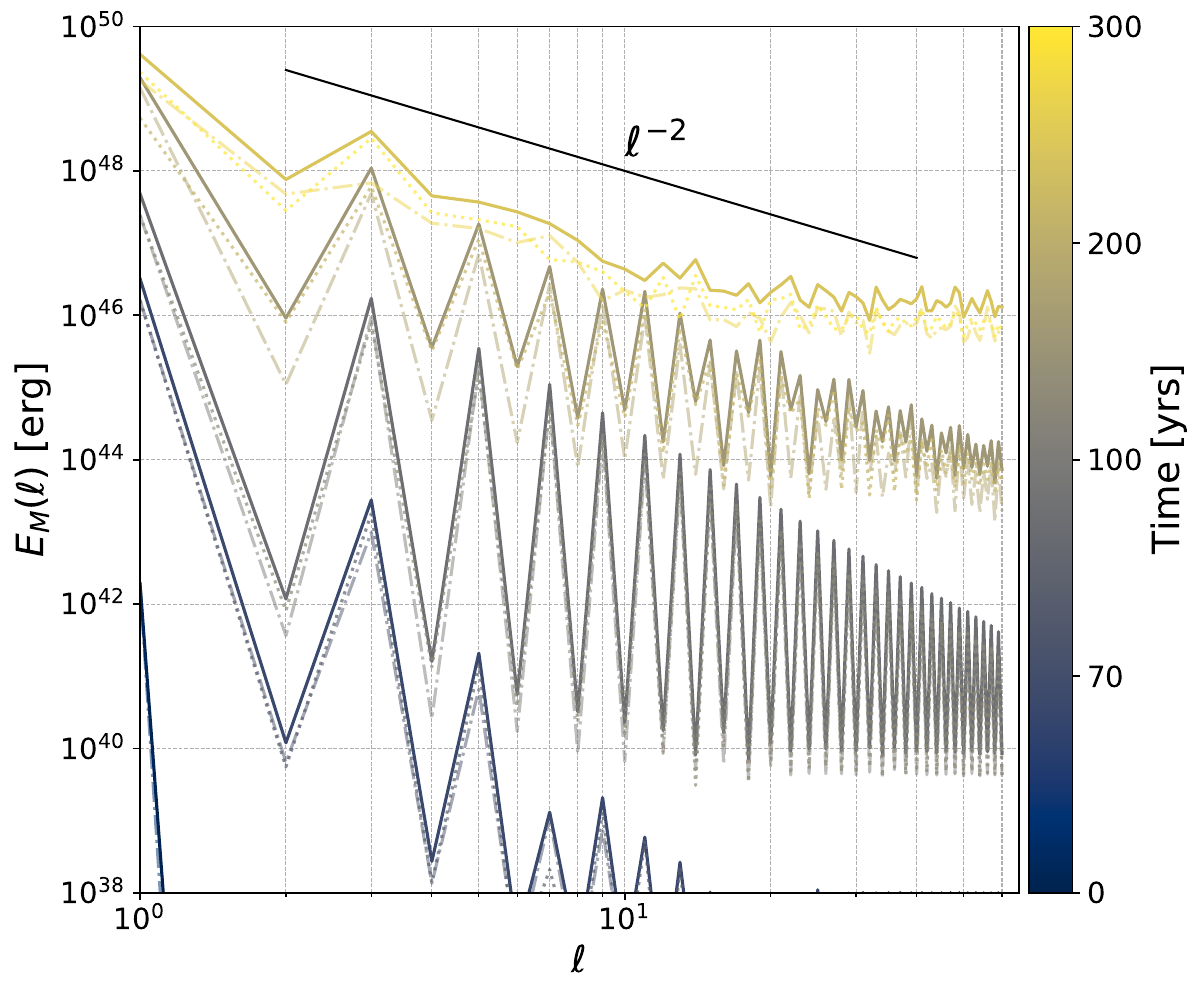}
    \caption{Spectral magnetic energy distribution: Solid lines show the total energy, dash-dotted lines depict the poloidal component, and dotted lines represent the toroidal contribution. Results are presented for $t=0$, $70$, $100$, $200$, and $300$\,yrs.}
    \label{fig: spectral energy}
\end{figure}

The simulation begins with a pulsar that has a dipolar field of approximately $10^{12}$\,G and an average magnetic field in the crust of a few $10^{12}$\,G. We adopt a field geometry primarily consisting of a dipolar poloidal field with no toroidal field to investigate the impact of CMI in generating helicity and, more specifically, dipolar toroidal fields. This contrasts with the Hall-MHD scenario, which can only produce even multipolar toroidal components from a pure dipole poloidal field. The question of how to create large-scale dipolar toroidal fields remains one of the unresolved challenges in most dynamo models.

Results are illustrated in Fig.\,\ref{fig: poloidal toroidal energy}, showing the time evolution of total magnetic energy, poloidal energy, and toroidal energy. We clearly distinguish three separate regimes. In the first stage, a linear growth of the toroidal field is observed, with the toroidal energy increasing proportionally to $t^2$, while the seed poloidal field remains constant. The second stage begins after around 50 years, once the strengths of the toroidal and poloidal magnetic fields become comparable. During this phase, the CMI induces an exponential growth of both components, progressing in near equipartition. Specifically, the magnetic energy grows as $\propto \mathrm{exp(0.3 t)}$ (red solid line), while the magnetic field grows as $\propto \mathrm{exp(0.15 t)}$, resulting in an average magnetic field of approximately $10^{15}$\,G in $\lesssim 100$\,yrs. Once the magnetic field is of the order of $10^{15}$\,G (about 100 years of NS's life), a third stage begins. Here, the non-linear Hall term (last term in \Eq{eq: electric field}) becomes important, altering the exponential growth. The combined effects of the Hall drift and the CMI lead to the magnetic field saturation at approximately a few $10^{16}$\,G after about 200 years. During this third stage, energy transfer between the poloidal and toroidal components leads to the typical oscillations between these field components, as observed in Hall-MHD simulations \cite{pons2007}.

In Fig.~\ref{fig: spectral energy}, we present the total energy spectrum (solid lines) along with the poloidal (dash-dotted lines) and toroidal (dotted lines) parts of the energy spectrum at various stages of evolution. After reaching equipartition during the second stage, as shown by the spectra at $t=70$ and $100$\,yrs, the magnetic energy is evenly distributed between poloidal and toroidal components across all multiples. Notably, the dipolar component exhibits significant growth, concentrating approximately 95\% of the magnetic energy within it. During the exponential growth stage, even multipoles contribute minimally to the total magnetic energy because the initial model was a pure dipolar field, and the CMI preserves the parity (odd or even nature) of modes during the evolution, analogous to the alpha dynamo effect \cite{raedler1980}. 
This behavior contrasts with the typical Hall-MHD scenario, where a dipolar poloidal field transfers most of its energy to a quadrupolar toroidal field, and the equipartition of magnetic energy occurs only in small-scales structures \cite{geppert1991,wiebicke1991,pons2007,dehman2023}.
Once the field has grown enough and the Hall term becomes important ($t > 200$ yrs), there is a significant transfer of energy from larger to smaller scales and between odd and even multipoles, leading to the characteristic $l^{-2}$ slope typical of Hall-MHD simulations. 

We stress again that, due to numerical limitations that prevent us from resolving centimeter scales, the growth time of the instability can be even faster for smaller scales, resulting in a turbulent-like field. Nevertheless, our main focus here is to propose a viable mechanism to generate large-scale strong fields and we believe that resolving smaller scales does not change our conclusions about the growth of the $l=1$ mode.
Interestingly, the magnetic field naturally saturates at magnetar strengths without any additional requirements. The exponential growth phase is initially driven by the $\EE \cdot \BB$ term in \Eq{eq: n5} until around 100 to 200 years, when the Hall drift term sets in. At late times, for $t \gtrsim 200$\,yrs, we enter a fully non-linear regime where the combined action of the Hall drift and the CMI limit the maximum
field strength. This behavior differs from CMI studies on the evolution of proto-NSs, where the saturation of the magnetic field is only attributed to the $\EE \cdot \BB$ term \cite{sigl2016}.

Furthermore, the CMI will also stop if the NS reaches exact $\beta$-equilibrium faster than the growth time of the instability. To further reinforce these results and provide a more precise quantitative prediction, it would be necessary to solve the coupled time evolution equation for $\delta \mu_5$ consistently within the assumed scenario (e.g., gravitational contraction due to spin-down, post-supernova fallback accretion, etc.) which is out of the scope of this work. In this letter, we aimed to highlight the possibility of a new plausible scenario for the origin of magnetar field strengths, leaving a more detailed investigation for future studies. 
In any event, we stress that the total chemical energy transferred to magnetic energy throughout the process is of the order of $10^{49}$\,erg, which is relatively small compared to the star's gravitational energy ($10^{53}$\,erg) or the typical rotational energy of a pulsar with a period of a few milliseconds ($10^{51-52}$\,erg). 
The CMI scenario only requires a residual contraction of about one meter to sustain a tiny $\delta \mu_5$ for several tens or hundreds of years. This is a fraction of the rotational energy loss employed to compress the crust enough to maintain matter slightly out of $\beta$-equilibrium.  

We have also explored the influence of the assumed deviation from chemical equilibrium on the growth rate of the magnetic field. In the model presented, with $\delta \mu /k_b T \approx 10^{-2}$, magnetar strength is achieved in about 75 years. During this period, the temperature remains relatively stable at approximately $10^9$\,K. We also ran a model with $\delta \mu /k_b T \approx 1$, yielding qualitatively similar results, but the magnetic field increased to $10^{15}$\,G in only 22 years. In contrast, smaller $\delta \mu$ values slow down the process. For instance, with $\delta \mu / k_b T \approx 10^{-4}$, it takes about 400 years to reach the same field strength, while with $\delta \mu / k_b T \approx 10^{-5}$, it takes about 6000 years. Therefore, we expect a significant growth of the magnetic field on timescales that are relatively short in an astrophysical context. Observed magnetars are aged between hundreds and tens of thousands of years, allowing sufficient time for the CMI to operate as long as $\delta \mu /k_b T \gtrsim 10^{-5}$. 
We have also confirmed that starting from different seed fields of a few Gauss, does not change qualitatively the final outcome. We observe exponential growth at the same rate and saturation occurs at roughly the same maximum strength.

A final noteworthy point concerns the geometry of the generated field. In many previous studies \cite{Geppert2012,Vigano2013,Gourg2019,dehman2020,deGrandis2021,dehman2023,dehman2023b} examining long-term magneto-thermal evolution and comparing it to observations, a crust-confined magnetic field with a strong large-scale toroidal component has been found to better match the observational phenomenology. However, the challenge has been in determining how to create such a field in the first place. Traditional dynamo mechanisms in the proto-NS phase can generate strong turbulent fields at very small scales but have not succeeded in producing a strong, large-scale toroidal dipole concentrated in the crust. In this work, we propose a mechanism that naturally leads to the type of geometry that aligns simulations with observations. This promising new approach, though not without its challenges, deserves further exploration in future research.

\begin{acknowledgments}
CD thanks Nordita for their hospitality during the summer of 2024. We also thank Axel Brandenburg, Dhrubaditya Mitra, and Arthur Suvorov for useful discussions. 
We acknowledge support from the Conselleria d'Educació, Cultura, Universitats i Ocupació de la Generalitat Valenciana, through grants CIPROM/2022/13, and ASFAE/2022/026 (with funding from
NextGenerationEU PRTR-C17.I1), and from the AEI grant PID2021-127495NB-I00
funded by MCIN/AEI/10.13039/501100011033. 
\end{acknowledgments}
\bibliography{apssamp}
\bibliographystyle{apsrev4-1}

\end{document}